

\input jnl
\input reforder

\rightline {NSF-ITP-92-85}
\rightline {YCTP-P17-92}
\rightline {CMU-HEP92-14}
\rightline {VAND-TH-92-9}
\rightline {June, 1992}

\title {SEMICLASSICAL GRAVITY AND INVISIBLE AXIONS}

\author { R. Holman${}^{a,b}$, T.W. Kephart${}^{a,c}$ \&
Soo-Jong Rey${}^{a,d,*}$}

\affil {
Institute for Theoretical Physics,
University of California, Santa Barbara CA 93106 ${}^a$
Physics Department, Carnegie Mellon University, Pittsburgh PA 15213 ${}^b$
Physics \& Astronomy Dept., Vanderbilt University, Nashville TN 37235${}^c$
Center for Theoretical Physics, Yale University, New Haven CT 06511 ${}^d$
}

\abstract
We show that charged Eguchi-Hanson instantons provide a
\sl concrete \rm and \sl calculable \rm new source of
intrinsic Peccei-Quinn symmetry breaking by quantum gravity.
The size of this breaking is shown to depend sensitively on
the short-distance details of a given theory,
but is generically suppressed by fermion zero modes.
Demanding that these gravitational effects not affect
the axion solution to the strong CP problem,
we find that at least two sets of quarks with differing
Peccei-Quinn charges are required.
In addition, these effects obviate the cosmological axion
domain wall problem but leave unchanged
problems associated with coherent axion oscillations.

\vskip1cm
\centerline {\sl submitted to Physical Review Letters}


\noindent *Yale-Brookhaven SSC Fellow. bitnet: Soo@Yalehep

\endtopmatter

The invisible axion\refto{invisible} was introduced more than
a decade ago as a phenomenologically viable solution to the
weak `strong-CP' problem.
The associated Peccei-Quinn (PQ) symmetry\refto{pq} is an exact
symmetry to all orders in perturbation theory,
but is broken intrinsically by nonperturbative QCD effects.
This breaking induces a mass of order $m_a \approx m_\pi
f_\pi / f_a$ for the invisible axion.
At temperatures larger than $\Lambda_{\rm QCD}$,
it is known that the axion mass diminishes rapidly\refto{gpy}.

It has always been assumed that the low energy physics of the
axion was dictated by the chiral symmetries of the light quarks.
Thus what was happening at high energies (other than the
spontaneous breaking of the PQ symmetry) appears to be essentially
irrelevant in determining low-energy axion dynamics.
This assumption, however, rests heavily on having the colored
matter content be as in the minimal standard model, e.g. having QCD
be asymptotically free.
In this case, only instantons of size $\sim \Lambda_{\rm QCD}^{-1}$
contribute to the axion potential.
It has been pointed out\refto{holdom, dineseiberg, flynnrandall}
that if extra colored particles are present in the theory above
the electroweak scale,
the contribution of small QCD instantons to axion physics cannot be
neglected.
Indeed in some situations, it is, in fact, the dominant contribution
to the axion potential.
This example clearly shows that short-distance physics can indeed
have a significant effect on axion physics.
We are then led to contemplate if and how non-QCD physics,
e.g. semiclassical gravity near the \sl Planck scale, \rm can
alter the axion physics in a similar way.

Since gravitational interactions are CP conserving,
perturbative quantum gravity should leave the PQ symmetry and
axion physics intact.
On the other hand, the CP symmetry could conceivably be
affected by {\it nonperturbative} quantum gravity effects.
Indeed, this possibility was recently pointed out in
Ref.\cite{pqgrav}, where the effects of higher dimension operators
which break the PQ symmetry (possibly induced by the exchange of
virtual black holes or by wormhole physics\refto{wormholes})
were investigated.
However, in Ref.\cite{pqgrav}, only plausibility arguments for
the \sl existence \rm of PQ symmetry breaking operators,
based on the classical black hole no-hair theorem and wormhole
physics were given.
There, it was impossible to estimate the \sl size \rm of PQ
symmetry breaking operators reliably.
Thus, the issue whether and how the PQ symmetry and axion physics
are modified by non-QCD short-distance physics remains unsettled.

In this Letter, we attempt to provide a more concrete analysis
of the question of whether Planck-scale physics will disturb the
PQ mechanism and low energy axion physics.
We do this by making use of certain well-known and well-studied
self-dual gravitational instantons\refto {gh}.
We assume they saturate the Euclidean path integral,
and hence should be included in the partition function describing
low energy physics.
While there may be many other potential quantum gravity effects
that might give rise to PQ breaking effects,
the approximation we use has the benefit of being both
well-motivated and above all, {\it controllably calculable}.

The relevant interactions to our discussion is described
by the action:
$$
\eqalign{
              S_0 = S_{\rm grav} & + S_{\rm gauge} +
                    S_{\rm fermion} + S_\theta \cr
   S_{\rm grav} & = \int \! d^4 x {\sqrt g}
                     [ -{M_{pl}^2 \over 16 \pi^2} R]
                     + \oint \! dn [K - K_0] \cr
  S_{\rm gauge} & = \int \! d^4 x {\sqrt g}
                   [{1 \over 4e^2} (F_{\mu \nu})^2 +
                   {1 \over 2 g^2}{\rm Tr}({\bf G}_{\mu \nu})^2]
                   \cr
S_{\rm ferm} & = \int \! d^4 x {\sqrt g} [\sum_{a=1}^{N_f}
                 \bar \psi_a \gamma^a e^\mu_a
                 (\nabla_\mu+Q^a_{\rm em}A_\mu +
                 {\bf B}_\mu(R^a)) \psi_a ]\cr
    S_\theta & = \int \! d^4 x {\sqrt g} [ { i \theta_{\rm QCD}
                 \over 16 \pi^2} Tr ({\bf G}_{\mu \nu} \tilde
                 {\bf G}_{\mu \nu})
               + {i \theta_{\rm em} \over 32 \pi^2}
                  F_{\mu \nu} \tilde F_{\mu \nu} +
                      { i \theta_{\rm grav}
                \over 384 \pi^2} (R \tilde R)]. }
\eqno (1)
$$
Here, $A_{\mu}$ and ${\bf B}_{\mu}$ comprise the electomagnetic
and gluon gauge fields,
$K$ the trace of the extrinsic curvature,
$e^{\mu}_a$ the vierbein, and
$\nabla_\mu = \partial_\mu + \omega_{\mu}$ the covariant
derivative for gravity.
This action supports the existence of not only the usual
QCD-Einstein instanton but also a self-dual configuration in both
the gravitational curvature
${\bf R}_{\mu \nu} \equiv R_{\mu \alpha \nu \beta} \Sigma^{\alpha
\beta}$
and the electromagnetic field strength (or that of any other
abelian (sub)group in the theory):
${\bf R}_{\mu \nu} = \pm \tilde {\bf R}_{\mu \nu}$
and $F_{\mu \nu} = \pm \tilde F_{\mu \nu}$.\refto{gh, eh}
The simplest such configuration was known as the Eguchi-Hanson
(EH) instanton\refto{eh}.
However, as we will see later, an instanton with nonzero
$F_{\mu \nu}$ turns out to be the most relevant to our
subsequent discussion of axion physics.
We will call these configurations Abelian Eguchi-Hanson (AEH)
instantons.
The explicit solution with core size
$\rho$ (where $x^2 \equiv x^\mu x^\mu \ge \rho^2$) is\refto{eh}:
$$
\eqalign{
           g^{\mu \nu} = & \,\, \delta^{\mu \nu}
                       - {\rho^4 \over x^4}
                         {x^\mu x^\nu \over x^2}
                       + {\rho^4 \over x^4 - \rho^4}
                         {\tilde x^\mu \tilde x^\nu \over x^2},
                       \cr
                 A_\mu = & \,\,\,
                         {2 P \rho^2 \tilde x_\mu \over x^4}.}
\eqno (2)
$$
The geometry of the manifold described by the above metric is
such that $\tilde x^\mu \equiv (y, -x, t, -z)$ is equivalent to
$x^\mu$ due to ${\bf R}\!{\rm P}_3$ global topology of the instanton.
The instanton $U(1)$ charges $P$ are to be chosen compatible with
the existence of spin structures.

The action of the AEH instanton is nonzero and given by
$$\eqalign{
         S_{AEH} = & \,\,\,{1 \over 4 e^2} \int d^4 x {\sqrt {-g}}
                   F_{\mu \nu}^2  \cr
                 = & \,\,\,{4 \pi^2 P^2 \over e^2}
                 = {\pi P^2 \over \alpha_e(\rho)}.}
\eqno (3)
$$
Note that the action is independent of Newton's constant, and
rather similar to that of the Yang-Mills instanton except for
a factor of two.
In fact, the analogy with Yang-Mills
instanton goes much deeper as we will see.
Consider a Dirac fermion $\psi$ in Eq.(1) of electric charge $Q$
and color representation $R$.
The axial current
$J_5^\mu \equiv \bar \psi \gamma_5 \gamma^\mu \psi$
is anomalous:
$$
\nabla^\mu J_{\mu 5} = {2 N_c Q^2_{\rm em} \over 16\pi^2}
                       F_{\mu \nu} \tilde F_{\mu \nu}
                     + {2 C_2 (R) \over 8 \pi^2} {\rm Tr}
                       {\bf G}_{\mu \nu} \tilde {\bf G}_{\mu \nu}
                     - {2 N_c \over 384 \pi^2} {\rm Tr}
                       {\bf R}_{\mu \nu} \tilde {\bf R}_{\mu \nu}.
\eqno (4)
$$
This implies a chiral charge asymmetry in the background of
an AEH-instanton \refto{thooft}:
$$\eqalign {
\Delta Q_5 (AEH) = \int [F \tilde F - {\rm Tr} {\bf R} \tilde {\bf R}]
= \cases{2  P^2 Q^2_{\rm em}, &if $P$ is an integer \cr
         2 (P^2 Q^2_{\rm em} - {1 \over 4}),& if $P$ is not an integer,}}
\eqno (5)
$$
i.e. there is {\it no} axial charge asymmetry contribution
from the purely gravitational sector,
but only from the $U_{\rm em}(1)$ field.
The $-{1 \over 4}$ contribution to the axial charge asymmetry in
case $P$ is not an integer is due to different boundary conditions
fermions have to satisfy across the antipodal points of the
identified spacetime described by the EH instanton.
Since the minimal electric charge of the standard model fermions
is $-1/3$,
the instanton charge should be restricted to $P \in 3 {\bf Z}$.

We now analyze the effects of AEH instantons on PQ symmetry breaking
when an invisible axion is introduced.
For definiteness, we consider an `extended' invisible axion model,
with isosinglet heavy quarks transforming under the representation $R$
of color and carrying nonzero PQ charges.
The size of the QCD induced axion potential is estimated to be:
$$ \eqalign{
      V_{\rm QCD} \approx & 2 K_{\rm QCD} \,
            \cos [N_{\rm QCD} {a \over f_a} + \theta_{\rm QCD}] \cr
            & K_{\rm QCD} =
            ({1 \over m_u} + {1 \over m_d} + {1 \over m_s})^{-1}
            \cdot <\!q \bar q\!>.}
\eqno (6)
$$
The integer $N_{\rm QCD}$ counts the QCD vacuum multiplicity
and is given by
$$
N_{\rm QCD} = {{2 \pi}\over{T_{\rm QCD}}}(2 {\rm Tr} [Q_{\rm pq}
{\bf T}_{a}^2 (R)]).
\eqno(7)
$$
Here $T_{\rm QCD}$ is the periodicity of the QCD theta parameter
\refto{sikivie, srednicki}, $Q_{\rm pq}$ denotes the PQ charge operator,
${\bf T}_{a}$ is a QCD color generator,
and the trace taken over all fermions.

AEH instantons of arbitrary neck size $\rho$ are all equally good
semiclassical solutions.
Therefore, in the semiclassical approximation to the path integral,
we need to sum over all instanton sizes $\rho \ge M_{\rm pl}^{-1}$.
Although a precise form of the measure $d\mu[\rho]$ is not known,
let us assume $d \mu [\rho] = d \ln \rho$ as in Yang-Mills case.

As is evident from Eq.(5), the $P=0$ EH instanton does not give
rise to an intrinsic breaking of PQ symmetry,
neither does a nontrivial axion potential.
The leading PQ breaking effect comes from the $P=3$ AEH instanton.
It produces $18 Q^2_{\rm em} $ chiral zero modes for each fermion,
and gives rise to an induced local operator of the form
$$  \eqalign{
{\cal O}_{AEH} [\rho] = & {1 \over \rho^4}
Det [\{(\bar E_R E_L)^9 \,
\prod_{\rm color} (\bar U_R U_L)^4 (\bar D_R D_L)\}^{N_g}
\prod_{\rm heavy} (\bar Q_R Q_L)^{9 Q_{\rm em}^2}] \cr
& \cdot \exp [-{9 \pi \over \alpha_e (\rho)}
+ i N_{\rm AEH} {a \over f_a} + i \theta_{\rm em}]. }
\eqno (8)
$$
Here, $N_g$ is the number of generations,
while $N_{\rm AEH} \equiv 9 \, N_c \, \sum_{\rm quark}
Q_{\rm pq} Q_{\rm em}^2 + 9 \sum_{\rm lepton} Q_{\rm pq}
Q_{\rm em}^2$.
$Det$ denotes the totally antisymmetrized, color, isospin and
hypercharge singlet products of quark and charged lepton zero modes.
The AEH instanton induced operator is strikingly
similar to that induced by QCD instantons.

In order to calculate the axion potential at low-energies,
we tie up the fermion zero modes via Yukawa interactions
with the Higgs fields.
Thus,
$ \bar \psi^i_L \psi^i_R \rightarrow \lambda_i H_{1,2}\,\Phi$
in which $\lambda$, $H_{1,2}$ and $\Phi$ denote specific Yukawa
coupling constants, and
the weak isodoublet and isosinglet Higgs fields, respectively.
{}From each Higgs loop integral in a size $\rho$ instanton background
and the fact that $<\!\! \Phi \!\!> \approx f_a$,
we get factors of $\lambda / 2 \pi$ and $2 \pi / \rho f_a$ for
$\rho \le f_a$ or $\rho f_a /2 \pi$ for $\rho \ge f_a$
respectively.
A more precise estimate of these factors and loop integrals can
be done following Dine and Seiberg\refto{dineseiberg} and
Flynn and Randall\refto{flynnrandall} (The possibility of increasing
the axion mass using small QCD instantons was originally suggested
by Holdom and Peskin\refto{holdom}).
Tying up the fermion zero modes through Yukawa couplings,
we find a new contribution to the axion potential due to $P=3$
AEH instantons:
$$ \eqalign {
V_{\rm AEH} [a] \approx & 2 K_{\rm AEH} \cos (N_{\rm AEH}
                          {a \over f_a} + \theta_{\rm em}), \cr
 K_{\rm AEH} \,\, = \, & ({9 \pi \over \alpha_e (\mu)})^{5 \over 2}
                 \, e^{-{9 \pi /\alpha_e (\mu)}}
                 \cases{ \eqalign { & \int_{M_{\rm pl}^{-1}}^{f_a^{-1}}
                     {d \rho \over \rho^5} (\mu \rho)^{b_0} \,
                     \prod [\lambda { f_a \rho \over 4 \pi^2}] \cr
                     + & \int_{f_a^{-1}}^{M_w^{-1}}
                     {d \rho \over \rho^5} (\mu \rho)^{b_0}
                     \prod [{ \lambda \over f_a \rho}]. }} }
\eqno (9)
$$
Here, $b_0$ denotes the first coefficient of the $\alpha_e$ beta function,
$\mu$ the renormalization scale (the inclusion of which renders the result
renormalization group invariant).
The largest contribution to the instanton size integrations come from
$\rho \approx f_a^{-1}$ for both terms in $K_{\rm AEH}$.
The new AEH instanton contribution is thus very sensitive to the matter
content of the theory near and above the PQ scale!

For illustration, let us take a model in which four color triplet
heavy quarks with electric charge $Q_{\rm em} = 2/3$ and mass
$10^{10} \, GeV$ are introduced.
Taking $\mu \approx f_a$, Eq.(9) gives
$$
{K_{\rm AEH} \over K_{\rm QCD}}
\approx 2 \times 10^{-10}
\, [{e^{-{9 \pi /\alpha_e (f_a)}} \over e^{-90 \pi}}] \,
[{m_Q \over 10^{10} \, GeV}]^{9 Q^2_{\rm em}} \,
[{f_a \over 10^{12} \, GeV}]^2.
\eqno (10)
$$
We have normalized the product of the light quark and lepton masses
to the standard model values with $m_t \sim 150 \,$ GeV, and
$\alpha_e (f_a)$ to 1/10 (as occurs in many supersymmetric models
containing a number of heavy charged scalar fields).
With these assumptions, we find that the new small instanton
induced potential is about one billionth of the QCD instanton
induced potential!
Our estimate is quite conservative, and varying the gauge and Yukawa
coupling constants certainly can and will make the size of this
contribution vary over a wide range.
As we will see later,
the estimate in Eq.(10) with $\alpha_e$ and $m_t$ as above
is roughly the upper bound to any small
instanton contributions to the axion potential based on current
experimental bounds on the neutron electric dipole moment(NEDM).

We first discuss how the invisible axion couples
to hadrons and photons at low energy.\refto{srednicki}
Below the QCD chiral symmetry breaking scale, the
low-energy dynamics of the Goldstone bosons, the invisible axion
and the photon is described by an effective Lagrangian.
Let us realize the $SU_L (3) \times SU_R (3) \times U_A (1)$
chiral symmetry nonlinearly using
$\Sigma \equiv \exp ({2 i \over f_\pi}
({\bf \Pi} + {\pi^0 \over \sqrt 6}{\bf I})) \in U(3)$ and let
$M = {\rm diag} (m_u, m_d, m_s)$ represent the current quark
mass matrix.
Then, assuming $f_\pi = f_{\eta'}$ for simplicity,
the low-energy chiral Lagrangian reads:
$$            \eqalign{
L_{\rm chiral} = & {1 \over 4} F_{\mu \nu} F^{\mu \nu} +
{1 \over 4} G_{\mu \nu}^a G^{\mu \nu a} +
{1 \over 4} f_\pi^2 {\rm Tr}
(\partial \Sigma^\dagger \partial \Sigma)
+ {1 \over 2} (\partial a)^2 + \cdots \cr
+ & <\! q \bar q\!> {\rm Tr} (M \Sigma^\dagger + h.c.) \cr
+ & \, {\rm Tr} [{\bf I}
\ln( \Sigma \, e^{i a /f_a + i \theta_{\rm QCD} })] \,
{1 \over 32 \pi^2} G_{\mu \nu}^a \tilde G_{\mu \nu}^a   \cr
+ & \, {\rm Tr} [{\bf Q}^2 \ln (\Sigma \, e^{i a / f_a +
i \theta_{\rm em}}) ] \,
{ 1 \over 32 \pi^2} F_{\mu \nu} \tilde F_{\mu \nu}.}
\eqno (11)
$$
where the ellipses denote higher derivative terms in
$\Sigma$ and $a$,
${\bf Q}_{\rm em} \equiv diag. \, (2/3, -1/3, -1/3)$
is the electric charge matrix and the traces are taken
over flavor space.
The last two terms come from the Wess-Zumino term which arises
after integrating out the quarks coupled to gluons and photon.
It ensures the correct anomalous $U_A (1)$ Ward identity Eq.(4)
involving both $U_{\rm em} (1)$ and QCD color anomalies.
For simplicity, we have ignored additional axion-photon
interactions, e.g. those coming from nonzero current quark masses.

Integrating out the QCD and AEH instanton fluctuations (in the
dilute gas approximation), the last two terms in Eq.(11) give:
$$  \eqalign{
L_{\rm inst} \approx &
K_{\rm QCD} \exp ({\rm Tr} ( {\bf I} \ln \Sigma) +
i N_{\rm QCD} {a \over f_a}  + i \theta_{\rm QCD})  \cr
+ & K_{\rm AEH} \exp ({\rm Tr} (9 {\bf Q}^2_{\rm em} \ln \Sigma )
+ i N_{\rm AEH} {a \over f_a} + i \theta_{\rm em}) \cr
+ & h.c.}
\eqno (12)
$$
The QCD and AEH instanton amplitudes $K_{\rm QCD}$ and
$K_{\rm AEH}$ are given as in Eqs.(6) and (9),
and the PQ charge weights are denoted by $N_{\rm QCD}$
and $N_{\rm AEH}$ respectively.

Having two independent sources for the axion potential,
one from small AEH instantons and the other from large QCD instantons,
the axion field may not settle to $-\theta_{\rm QCD}$
at the hadronic scale.
{}From Eqs.(6) and (12), and ignoring a small mixing
($\sim {\cal O}(f_\pi / f_a)$) between the $\eta'$ and the axion,
we find the new axion minima at:
$$\eqalign{
\bar \theta_{\rm QCD} & \equiv N_{\rm QCD} {<\! a\!> \over f_a}
                      + \theta_{\rm QCD}     \cr
                      & = -N_{\rm AEH} \cdot
                       {N_{\rm AEH} \over N_{\rm QCD}}
                       \cdot {K_{\rm AEH} \over K_{\rm QCD}}
                       \cdot ({\theta_{\rm QCD} \over N_{\rm QCD}}
                     - {\theta_{\rm em} \over N_{\rm AEH}} ).}
\eqno (13)
$$
Typically, we expect
$\Delta \theta \equiv \theta_{\rm QCD} - \theta_{\rm em} \approx O(\pi)$
and $N_{\rm QCD} \approx N_{\rm AEH}$.
Then, $\bar \theta_{\rm QCD} \approx {K_{\rm AEH} / K_{\rm QCD}}
\cdot \Delta \theta$.
Thus, demanding that $\bar{\theta}_{{\rm QCD}}$ be less than $10^{-9}$,
as required by NEDM measurements,
yields $K_{\rm AEH}/K_{\rm QCD} \le 10^{-10}$!
This bound is indeed met by Eq.(10) as long as not too many charged
and colored PQ charge carrying fermions are introduced at short distances.
The invisible axion still can solve the `strong-CP' problem.

The physical axion mass is obtained from Dashen's theorem:
$$
m_a^2 f_a^2 \approx [ Q_a, [Q_a, L_{\rm inst}]]
\eqno (14)
$$
in which $Q_a \equiv \int d^3 \vec x \, J_a^0$
is the physical axion charge operator.
Let us see how the AEH instanton induced potential modifies the
invisible axion and other Goldstone boson mass spectrum.
These modifications are obtained most easily by diagonalizing
the total Goldstone boson mass matrix from the second line of Eq.(11)
and QCD and AEH instanton induced potentials in Eq.(12).
{}From this we find the physical axion mass:
$$
m_a^2 f_a^2 \approx K_{\rm QCD}  + K_{\rm AEH} +
{\cal O} ({K^2_{\rm AEH} \over K_{\rm QCD}})
\eqno (15)
$$
In addition, we find that both the $\eta$ {\it and} the $\eta'$
receive mass corrections due to small AEH instantons proportional
to $K_{\rm AEH}/f_\pi^2$.

If the gravitational effects discussed above saturate the bounds
from NEDM measurements,
they can also solve the cosmological domain wall problem of axion
theories.\refto{sikivie}
Recall that domain walls arise due to the fact that a
$Z_{N_{\rm QCD}}$ subgroup of the $U(1)_{\rm pq}$ symmetry
may be anomaly free with respect to QCD,
but maybe broken spontaneously by both the Higgs vacuum
expectation values and the quark condensate.
We now repeat the same calculation for the gravity induced potential.
Using the fact that all fermions couple equally to gravity,
we can easily see that the subgroup of $U(1)_{\rm pq}$ free from
gravitational anomalies is $Z_{N_{\rm grav}}$ where
$N_{\rm grav}$ is defined by:
$N_{\rm grav} \equiv {\rm Tr} [Q_{\rm pq}]$.
Furthermore, $Z_{N_{\rm em}}$, where
$N_{\rm em} \equiv 9 {\rm Tr} [Q_{\rm pq} Q_{\rm em}^2]$
is the subgroup of $U(1)_{\rm pq}$ which is anomaly-free
with respect to the $U_{\rm em}(1)$.
The full axion potential takes the form:
$$
\eqalign{
V(a) & = 2\,K_{\rm QCD} \, \cos(N_{\rm QCD} {a \over f_a}
                         + \bar \theta_{\rm QCD} ) \cr
     & +\,2\,K_{\,\rm grav}\,\,\cos (N_{\,\rm grav}\,{a \over f_a}
                         + \bar  \theta_{\, \rm grav}) \cr
     & +\,\, 2\,\, K_{\,\rm em} \, \,\,\cos (\, N_{\, \rm em} \,\,
               {a \over f_a} \, + \, \bar \theta_{\,\rm em}\,\,).}
\eqno (16)
$$
As we argued earlier, for the matter content of the
standard model, the purely gravitational contribution
can be neglected compared to the other two.

It is clear that if $N_{\rm QCD}$ does not divide
$N_{\rm em}$, then the $N_{\rm QCD}$ vacua of pure QCD
which were originally degenerate will have their degeneracy
split by an amount $ \sim 2\, K_{\rm em}$.
For ${K_{\rm em} \over K_{\rm QCD}}  \sim 10^{-10}$,
this splitting is enough to bias the $N_{\rm QCD}$ multiple
vacua so that the lowest energy vacuum percolates.
Hence, new but tiny nonperturbative effects
due to small AEH instantons can solve the cosmological
domain wall problem.

Our analysis up to this point was made under the assumptions that
the matter content is that of a certain extended class of the
standard model and that the AEH instantons represent the
characteristic size of nonperturbative semiclassical gravity effects.
It should always be kept in mind that these assumptions may not hold.
In such cases, the axion solution to `strong CP' problem may be
endangered by short-distance physics.
If this occurs, the axion solution can still be saved by decoupling
the fermion PQ current from the gravity and $U(1)$ gauge field,
but not from QCD.
This implies
$$ \eqalign{
& {\rm Tr}\, [Q_{\rm pq} {\bf T}_a^2]  \ne 0,  \cr
& {\rm Tr}\, [Q_{\rm pq} {Q_{\rm em}^2}] = 0, \cr
& {\rm Tr}\, [Q_{\rm pq}] = 0.}
\eqno (17)
$$
These require, in general, that the quarks must be at least
in two different color and $U_{\rm em} (1)$ representations.
We note that a similar observation was made by Georgi and Wise
\refto{georgiwise} in the context of grand unified theory axion models.
To be free of cosmological domain wall problem, $N_{\rm QCD} = 1$,
or inflation is also required.

Since size of the gravity-induced axion potential is rather small,
there is virtually no change in the cosmological coherent axion
oscillation and associated axion energy density problem.
Since the new contribution to the axion potential by small instantons
is most pronounced near $f_a$,
finite-temperature effects are insignificant at the time of potential
coherent oscillations.
Even if the new induced axion potential were sizeable,
e.g. $\theta_{\rm QCD} \approx \theta_{\rm AEH}$ and
$K_{\rm AEH} >\!> K_{\rm QCD}$,
one would still have to overcome finite-temperature thermal damping.
The damping rate is larger than that of the universe expansion
as long as $T\ge 10^{4}$ GeV.\refto{mclerran}
Due to this overdamping, coherent axion oscillations will not start
any sooner than around the electroweak scale or $K_{\rm AEH} /f_a^2$,
whichever is larger.
Since the generated coherent axion energy is sufficiently redshifted
away by now,
the PQ scale $f_a$ is still bounded above by $\approx 10^{12}$ GeV.

Finally, a natural place for AEH instantons to appear is in
four-dimensional compactified superstring theories.\refto{rey92}
However, there are a few minor differences in that case.
First, there exists a nontrivial dilaton field which grows stronger
near the core of the instanton.
Thus, the above instanton sum should be modified correspondingly.
The final form is a nontrivial action for the zero mode of the dilaton
field.
Also, it is necessary to keep higher derivative terms in calculating
the instanton action.
Since the sign in front of curvature squared term is negative,
the resulting instanton action may actually be much smaller than
that in the pure Einstein theory discussed in the present paper.
These two aspects add together in a way to increase the total
axion potential size.
This will be discussed in detail in a separate paper\refto{rey92}.

To conclude, we have given a \sl concrete \rm  and \sl calculable \rm
realization of the possibility that short distance effects such as
semiclassical gravity might disturb the PQ mechanism.
Indeed, the very same argument goes through for any global symmetries
spontaneously broken at a relatively low energy scale.
It is clear that axion physics does generically depend sensitively on
the details of physics at short distance scales.
While we have looked at effects of short-distance physics such as
gravity on the axion,
these effects would manifest themselves equally strongly on low-energy
pseudo-Goldstone bosons in QCD:
for example, the mass of the $\eta$ and the $\eta'$ were seen to
depend sensitively on short-distance physics,
and hence may serve as a potential probe of QCD at short-distances.

SJR thanks M. Einhorn, L. McLerran, E. Mottola, M. Srednicki and
G. `t Hooft for useful discussions.
This work was completed at the Institute for Theoretical Physics at
Santa Barbara, who we thank for hospitality.
This work was supported in part by the National Science Foundation
under Grant No. PHY89-04035.
R.H. was supported in part by DOE grant DE-AC02-76ER3066,
T. W. K. was supported in part by DOE grant DE-FG05-85ER40226 and
S.J.R. was supported in part by funds from
Texas National Research Laboratory Commission.

\references
\refis {invisible} J.E. Kim, Phys. Rev. Lett. \bf 43  \rm (1979) 103;
M. Shifman, A.I. Vainshtein and V. Zakharov, Nucl. Phys. \bf B166 \rm
(1980) 4933; M. Dine, W. Fischler and M. Srednicki, Phys. Lett.
\bf B104 \rm (1981) 199; A. Zhitniskii, Yad. Fiz. \bf 31 \rm (1980)
497 [Sov. J. Nucl. Phys. \bf 31 \rm (1980) 260].

\refis {pq} R. Peccei and H.R. Quinn, Phys. Rev. Lett. \bf 38 \rm
(1977) 1440: Phys. Rev. \bf D16 \rm (1977) 1791.

\refis {pqgrav} R. Holman, S. Hsu, T. Kephart, E. Kolb, R. Watkins
and L. Widrow, Phys. Lett. \bf 282B \rm (1992) 132;
M. Kamionkowski and J. March-Russell, Phys. Lett. \bf 282B \rm (1992) 137;
S.M. Barr and D. Seckel, BA-92-11 preprint (1992);
S.M. Lusignoli and M. Roncadelli, FNT/T92/12 preprint (1992).

\refis {wormholes}
H.B. Nielsen and M. Ninomiya, Phys. Rev. Lett. \bf 62 \rm (1989) 1429;
K. Choi and R. Holman, Phys. Rev. Lett. \bf 62 \rm (1989) 2575 ;
J. Preskill, S. Trivedi and M.B. Wise,  Phys. Lett. \bf B223 \rm (1989)
26.

\refis {dineseiberg}  M. Dine and N. Seiberg, Nucl. Phys. \bf B273
\rm (1986) 109.

\refis {flynnrandall}
J. Flynn and L. Randall, Nucl. Phys. \bf B293 \rm (1987) 731.

\refis {sikivie} P. Sikivie, Phys. Rev. Lett. \bf 48 \rm (1982) 1156.

\refis {gpy} D.G. Gross, R. Pisarski and L.G. Yaffe, Rev. Mod. Phys.
\bf 53  \rm (1981) 43.

\refis {thooft} G. `t Hooft, Nucl. Phys. \bf B315 \rm (1989) 517.

\refis {eh} T. Eguchi and A. Hanson, Phys. Lett. \bf 74B \rm (1978) 430:
Ann. Phys. (NY) \bf 120 \rm (1979) 82.

\refis {gh} G.W. Gibbons and S.W. Hawking, Phys. Lett. \bf 78B \rm
(1978) 430.

\refis {rey92} S.-J. Rey, \sl Gravitational Superstring
Instantons and Solitons, \rm NSF-ITP-92-93 (1992).

\refis {holdom}B. Holdom and M.E. Peskin,  Nucl. Phys. \bf B208 \rm
(1982) 397; B. Holdom, Phys. Lett. \bf 154B \rm (1985) 316, ibid (E)
\bf 156B \rm (1985) 452.

\refis {mclerran} E. Mottola, L. McLerran and M.E. Shaposhnikov,
Phys. Rev. \bf D43 \rm (1991) 2027.

\refis {srednicki} M. Srednicki, Nucl. Phys. \bf B260 \rm (1985)
689.

\refis {georgiwise} H. Georgi and M.B. Wise, Phys. Lett. \bf  B \rm
(1981) 123.

\endreferences

\endit
\end